\documentclass[12pt]{article}

\usepackage[margin=1in]{geometry}
\usepackage[utf8]{inputenc}
\usepackage[T1]{fontenc}
\usepackage{amsmath,amssymb,bm,mathtools}
\usepackage{graphicx}
\usepackage{booktabs}
\usepackage{array}
\usepackage{multirow}
\usepackage{siunitx}
\usepackage{xcolor}
\usepackage{url}
\usepackage[colorlinks=true,linkcolor=blue,citecolor=blue,urlcolor=blue]{hyperref}
\usepackage{enumitem}
\usepackage{microtype}
\usepackage{subcaption}
\usepackage{float}

\newcommand{\CavityShape}{$48\times36\times30$ lattice nodes}
\newcommand{\CavityGeometry}{rectangular}
\newcommand{\SourceNodeCount}{952}
\newcommand{\DatasetCaseCount}{640}
\newcommand{\FrequencyCount}{16}
\newcommand{\FrequencyRange}{$0.00391$--$0.0235$ cycles per lattice time step}
\newcommand{\ReceiversPerCase}{512}
\newcommand{\InteriorCollocationCount}{256}
\newcommand{\RigidWallCollocationCount}{128}
\newcommand{\LBMTau}{$0.56$}
\newcommand{\LBMSteps}{7200}
\newcommand{\LBMFitStart}{4200}
\newcommand{\LBMRampSteps}{750}
\newcommand{\LatentRank}{96}
\newcommand{\HiddenWidth}{256}
\newcommand{\NetworkDepth}{5}
\newcommand{\CoordinateFourierBands}{6}
\newcommand{\FrequencyFourierBands}{7}
\newcommand{\BranchSensorCount}{256}
\newcommand{\TrainingEpochs}{200}
\newcommand{\TrainingBatchCases}{2}
\newcommand{\MaxReceiverPoints}{768}
\newcommand{\TrainingFraction}{80\%}
\newcommand{\ValidationFraction}{10\%}
\newcommand{\TestingFraction}{10\%}
\newcommand{\InitialLearningRate}{$2\times10^{-4}$}
\newcommand{\WeightDecay}{$10^{-6}$}
\newcommand{\RandomSeedCount}{five}

\newcommand{\ModeFrequencyError}{0.0325}
\newcommand{\ModeShapeError}{0.0625}
\newcommand{\ModeShapeCorrelation}{0.998}
\newcommand{\SolverLinearityError}{$1.59\times10^{-5}$}
\newcommand{\ConvergenceOrder}{0.549}
\newcommand{\PistonFrequencyRatio}{0.4}
\newcommand{\PistonComplexError}{0.182}
\newcommand{\PistonAmplitudeError}{0.115}
\newcommand{\PistonPhaseRMSE}{8.78}

\newcommand{\CLBORelativeError}{0.184}
\newcommand{\CLBORelativeErrorStd}{0.00771}
\newcommand{\DeepONetRelativeError}{0.367}
\newcommand{\DeepONetRelativeErrorStd}{0.00742}
\newcommand{\CLBOAmplitudeError}{2.32}
\newcommand{\DeepONetAmplitudeError}{3.78}
\newcommand{\CLBOPhaseError}{17.2}
\newcommand{\DeepONetPhaseError}{28.9}
\newcommand{\CLBOComplexCorrelation}{0.913}
\newcommand{\DeepONetComplexCorrelation}{0.841}

\newcommand{\CLBOLinearityError}{$1.31\times10^{-7}$}
\newcommand{\DeepONetLinearityError}{0.406}
\newcommand{\CLBOZeroInputError}{$0$}
\newcommand{\CLBOPermutationError}{$9.10\times10^{-8}$}
\newcommand{\CLBOMeshTransferError}{1.00}
\newcommand{\DeepONetMeshTransferError}{2.33}
\newcommand{\CLBOSourceFamilyError}{0.402}
\newcommand{\DeepONetSourceFamilyError}{0.917}

\newcommand{\UnseenCombinationCount}{5}
\newcommand{\CLBOUnseenCombinationError}{0.237}
\newcommand{\DeepONetUnseenCombinationError}{0.415}
\newcommand{\CLBOUnseenMedianError}{0.177}
\newcommand{\DeepONetUnseenMedianError}{0.474}
\newcommand{\CLBOUnseenComplexCorrelation}{0.961}
\newcommand{\DeepONetUnseenComplexCorrelation}{0.919}
\newcommand{\CLBOUnseenWinRate}{100\%}
\newcommand{\DeepONetUnseenWinRate}{0\%}
\newcommand{\UnseenPairedAdvantage}{0.178}
\newcommand{\UnseenAdvantageCILower}{0.117}
\newcommand{\UnseenAdvantageCIUpper}{0.249}

\newcommand{\ReceiverInterpolationCaseCount}{64}
\newcommand{\CLBOReceiverInterpolationMean}{0.332}
\newcommand{\CLBOReceiverInterpolationMedian}{0.245}
\newcommand{\DeepONetReceiverInterpolationMean}{0.581}
\newcommand{\DeepONetReceiverInterpolationMedian}{0.427}

\newcommand{\CLBOTenPercentError}{1.04}
\newcommand{\CLBOTwentyFivePercentError}{0.644}
\newcommand{\CLBOFiftyPercentError}{0.381}
\newcommand{\DeepONetTenPercentError}{1.00}
\newcommand{\DeepONetTwentyFivePercentError}{0.882}
\newcommand{\DeepONetFiftyPercentError}{0.626}
\newcommand{\NoQuadratureError}{0.307}
\newcommand{\PhysicsCLBOError}{0.183}
\newcommand{\CLBOInferenceTime}{14.5}
\newcommand{\DeepONetInferenceTime}{7.04}
\newcommand{\LBMCaseTime}{$2.65\times10^{5}$}
\newcommand{\CLBOSpeedup}{$1.83\times10^{4}$}
\newcommand{\DeepONetSpeedup}{$3.8\times10^{4}$}
\newcommand{\CLBOParameterCount}{653952}
\newcommand{\DeepONetParameterCount}{770432}
\newcommand{\HardwareDescription}{CPU laptop (WSL2), single precision (\texttt{float32}) inference without GPU acceleration}

\newcommand{\paperfigure}[3][1.0\textwidth]{%
  \IfFileExists{#2}{%
    \includegraphics[width=#1]{#2}%
  }{%
    \fbox{\parbox[c][5.4cm][c]{#1}{\centering
      \textbf{Figure file not yet generated}\\[0.6em]
      \texttt{\detokenize{#2}}\\[0.6em]
      #3}}
  }%
}

\newcommand{\clbo}{\textsc{clbo}}
\newcommand{\deeponet}{DeepONet}
\newcommand{\Aop}{\mathcal{A}}
\newcommand{\vhat}{\widehat{v}_{n}}
\newcommand{\phat}{\widehat{p}}

\begin{document}

\title{\vspace{-1.0cm}
\Large\bfseries
Quadrature-Aware Complex-Linear Neural Operator for\\
Boundary-to-Field Prediction in Resonant Acoustics
}

\author{%
Muhammad Idrees Khan\thanks{Corresponding author: Muhammad Idrees Khan}
\hspace{1.5em}
Hua-Dong Yao\\[0.45em]
\parbox{0.88\textwidth}{\centering
\normalsize Department of Mechanical Engineering, Chalmers University of Technology\\
\normalsize Gothenburg 41296, Sweden
}%
}

\date{}

\maketitle

\begin{abstract}
  Repeated prediction of acoustic fields from spatially distributed boundary excitation is computationally expensive when each source realization requires a new wave simulation. This work introduces a quadrature-aware complex-linear boundary operator (\clbo) that maps complex normal velocity on a vibrating surface to complex pressure at receiver locations. The model couples learned source and receiver basis functions through an explicit complex surface-quadrature contraction, so the boundary excitation enters linearly by construction. This preserves complex superposition, homogeneity, and zero response to zero excitation, while representing the source through coordinates, normals, and quadrature weights rather than a fixed flattened input vector. Reference data were generated using a verified three-dimensional multiple-relaxation-time (MRT) lattice Boltzmann solver and stored in a solver-agnostic boundary-to-field format. \clbo{} was compared with a fixed-sensor complex \deeponet{} under matched case splits and optimization settings, with additional tests of structural consistency, receiver-coordinate interpolation, source discretization, source-family holdout, label efficiency, physics-informed ablations, unseen source mixtures, and computational cost. Across \RandomSeedCount{} training seeds, \clbo{} achieved a mean complex relative field error of \CLBORelativeError{} $\pm$ \CLBORelativeErrorStd{}, compared with \DeepONetRelativeError{} $\pm$ \DeepONetRelativeErrorStd{} for \deeponet{}. Its measured source-superposition error was \CLBOLinearityError{}, and its mean error on newly simulated mixed-source cases was \CLBOUnseenCombinationError{}, compared with \DeepONetUnseenCombinationError{} for \deeponet{}. Inference was \CLBOSpeedup{} faster than the reference calculation for the reported query size. These results show that enforcing the known complex-linear boundary-to-field structure improves physical consistency and generalization under distributed acoustic excitation.
  \end{abstract}


\section{Introduction}

Many acoustic and vibroacoustic analyses require repeated evaluation of the pressure field generated by a spatially distributed vibrating boundary. Such boundary-to-field predictions are needed when assessing how changes in wall vibration, actuator patterns, source placement, operating frequency, or receiver location affect the resulting interior sound field. Once the geometry, medium properties, operating frequency, and passive boundary conditions are fixed, the acoustic stage is a complex-linear operator from prescribed normal velocity to pressure. Finite-element, boundary-element, modal, and time-domain wave solvers provide high-fidelity solutions, but repeated calculations over source fields, frequencies, and receiver sets can dominate design exploration, uncertainty studies, and surrogate-based optimization \cite{morse1968,kuttruff2016,ihlenburg1998,wu2000,james2006}.

Operator learning seeks to replace repeated numerical solves by learning maps between input and output functions. The Deep Operator Network (DeepONet) uses branch and trunk networks to represent an operator from input-function samples to coordinate queries \cite{lu2021}; neural-operator frameworks use learned integral kernels and include low-rank, graph, and Fourier parameterizations \cite{kovachki2023,li2021fno}; geometry-aware variants address irregular discretizations and varying domains \cite{li2023gino}. Physics-informed neural networks and physics-informed neural operators introduce governing-equation residuals into optimization \cite{raissi2019,li2021pino}, although such residual losses can introduce conditioning and loss-balancing difficulties \cite{krishnapriyan2021}.
More broadly, machine learning has become a common tool for data-driven modeling, closure modeling, reduced-order modeling, and control in fluid mechanics \cite{brunton2020mlfluid}.

Several related ideas are important for the present problem but do not by themselves enforce its complete structure. Sum aggregation and point-set architectures provide permutation-invariant representations of unordered inputs \cite{zaheer2017,qi2017pointnet,lee2019settransformer}; kernel neural operators use numerical quadrature for geometrically flexible function-space approximation \cite{lowery2024kno}; recent low-rank kernel models learn separable or singular-function representations \cite{koren2025svdno}; and boundary-integral neural methods reduce partial-differential-equation (PDE) learning to boundary quantities \cite{fang2023boundary,qu2023}. In acoustics, neural and physics-informed models have been used for parameterized propagation, continuous neural acoustic fields, sound-field reconstruction, room responses, radiation, and acoustic transfer \cite{borrel2021,borrel2024,luo2022naf,jin2022,ma2024,karakonstantis2024,jin2025nat}. These studies establish the promise of learned wave surrogates, but generic nonlinear source encoders do not in general guarantee exact complex superposition. Moreover, fixed-sensor vector encodings, such as the baseline used here, do not represent the source surface through explicit quadrature weights.

This work introduces a quadrature-aware complex-linear boundary operator, denoted \clbo. Learned nonlinear networks represent source- and receiver-dependent basis functions, while the complex source velocity enters only through a weighted surface contraction. The architecture is therefore nonlinear in coordinates and frequency but exactly complex-linear in the distributed source field. It is also invariant to consistent permutations of source samples and accepts a variable number of source quadrature points. The present contribution lies in integrating low-rank source--receiver bases, permutation-invariant surface aggregation, and explicit quadrature into a complex boundary-to-field surrogate whose source algebra follows linear acoustics by construction. The resulting formulation preserves the known dependence on prescribed boundary velocity while enabling direct tests of generalization to newly simulated complex source combinations. Recent boundary-integral and low-rank neural operators further motivate this structure, but address different PDEs, geometries, or training objectives \cite{koren2025svdno,qu2023}.

The contribution is evaluated in a resonant three-dimensional rectangular cavity using reference data generated by a multiple-relaxation-time (MRT) lattice Boltzmann method (LBM). The paper makes four contributions:

\begin{enumerate}[leftmargin=1.8em]
\item a low-rank boundary-to-field neural operator that preserves complex source superposition, homogeneity, and zero-input consistency by construction;
\item a surface representation using source coordinates, normals, and quadrature weights rather than a fixed flattened boundary vector;
\item a direct generalization test in which new complex source mixtures are simulated with MRT-LBM and evaluated against both learned models; and
\item a verified evaluation workflow from reference calculations to standardized cases, fixed case-level splits, training, evaluation, and ablation studies.
\end{enumerate}

The study concerns one fixed cavity and a linear time-harmonic acoustic regime in nondimensional lattice units. The data interface is solver-agnostic: another verified numerical method or measurement system can provide the labels when it supplies the same boundary and field quantities. A trained checkpoint nevertheless represents the geometries and passive boundary conditions covered by its training data unless those quantities are introduced explicitly as model inputs.

\section{Boundary-to-field problem}

\subsection{Frequency-domain acoustics}

Let $\Omega\subset\mathbb{R}^{3}$ be a bounded acoustic domain with boundary $\Gamma=\Gamma_v\cup\Gamma_r$, where $\Gamma_v$ is a vibrating surface and $\Gamma_r$ is rigid. With the convention
\begin{equation}
 p(\bm{x},t)=\Re\left\{\phat(\bm{x},\omega)e^{i\omega t}\right\},
\end{equation}
the complex pressure satisfies
\begin{align}
 \nabla^2\phat+k^2\phat&=0 && \text{in }\Omega, \\
 \frac{\partial\phat}{\partial n}&=-i\rho_0\omega\vhat && \text{on }\Gamma_v, \\
 \frac{\partial\phat}{\partial n}&=0 && \text{on }\Gamma_r,
 \label{eq:helmholtz_problem}
\end{align}

where $k=\omega/c_0$ and \(\vhat=\widehat{\bm{u}}\cdot\bm{n}\) denotes the prescribed normal velocity, positive in the outward-normal direction. Away from exact lossless cavity eigenfrequencies, or when the finite damping present in the reference calculation is included, Eq.~\eqref{eq:helmholtz_problem} defines a complex-linear map

\begin{equation}
 \Aop_{\omega}:\vhat(\bm{s})\longmapsto\phat(\bm{x},\omega).
 \label{eq:operator_definition}
\end{equation}
Consequently,
\begin{equation}
 \Aop_{\omega}(a v_1+b v_2)
 =a\Aop_{\omega}(v_1)+b\Aop_{\omega}(v_2),
 \qquad a,b\in\mathbb{C}.
 \label{eq:physical_linearity}
\end{equation}

\subsection{Discrete surface representation}

A sampled source surface is represented by
\begin{equation}
 \mathcal{S}_h=\left\{(\bm{s}_j,\bm{n}_j,A_j,v_j)\right\}_{j=1}^{N_s},
\end{equation}
where $\bm{s}_j$ is a source location, $\bm{n}_j$ is the associated normal, $A_j>0$ is a quadrature weight, and $v_j\in\mathbb{C}$ is the sampled normal velocity. A direct learned kernel would take the form
\begin{equation}
 \phat(\bm{x},\omega)
 \approx\sum_{j=1}^{N_s}A_j
 K_{\theta}(\bm{x},\bm{s}_j,\bm{n}_j,\omega)v_j,
 \label{eq:learned_kernel}
\end{equation}
with cost proportional to $N_sN_r$ for $N_r$ receiver queries. This expression is analogous to a learned boundary-integral or Kirchhoff--Helmholtz transfer representation, but with the Green-type kernel replaced by a trainable source--receiver kernel. Classical and modern acoustic boundary-element formulations motivate this boundary-only view while also highlighting the dense-kernel cost that the present low-rank factorization avoids \cite{wu2000,williams1999,wout2021bem}. The proposed model factorizes this interaction into source and receiver bases.

\section{Quadrature-aware complex-linear boundary operator}

\subsection{Coordinate and frequency features}

Coordinates and frequencies are normalized using statistics computed exclusively from the training split. The same feature map is used for source and receiver coordinates. For a normalized scalar coordinate $q$, Fourier features are
\begin{equation}
 \gamma(q)=\left[q,\sin(2^0\pi q),\cos(2^0\pi q),\ldots,
 \sin(2^{B-1}\pi q),\cos(2^{B-1}\pi q)\right].
\end{equation}
The implementation configuration used in this study uses \CoordinateFourierBands{} coordinate bands and \FrequencyFourierBands{} frequency bands. Fourier feature mappings are used to improve representation of oscillatory coordinate dependence \cite{tancik2020}. Complex quantities are stored as paired real and imaginary channels, while their products are evaluated with exact complex arithmetic implemented over those channels; this differs from using unconstrained complex-valued hidden layers \cite{trabelsi2018}.

\subsection{Low-rank source and receiver bases}

\begin{figure}[H]
\centering
\paperfigure{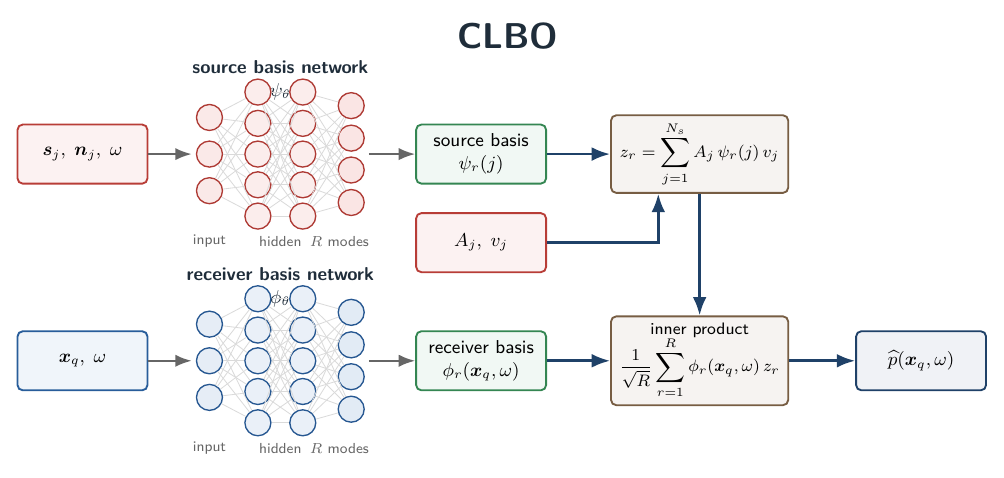}{Schematic of the quadrature-aware complex-linear boundary operator.}
\caption{Architecture of the proposed \clbo{}. A source basis network maps each source location, normal, and frequency to $R$ complex basis functions after normalized coordinate and frequency features. The prescribed boundary velocity enters only through the weighted surface sum in Eq.~\eqref{eq:latent_source}. A receiver basis network maps each query location and frequency to $R$ complex basis functions, and the predicted pressure follows from the low-rank inner product in Eq.~\eqref{eq:clbo_prediction}. The multilayer blocks are schematic; the implemented encoders use depth~\NetworkDepth{}, hidden width~\HiddenWidth{}, and latent rank~\LatentRank{}.}
\label{fig:clbo_architecture}
\end{figure}

The source network produces a complex basis
\begin{equation}
 \bm{\psi}_{\theta}(\bm{s}_j,\bm{n}_j,\omega)\in\mathbb{C}^{R},
\end{equation}
and the receiver network produces
\begin{equation}
 \bm{\phi}_{\theta}(\bm{x}_q,\omega)\in\mathbb{C}^{R}.
\end{equation}
The latent source coefficients are obtained by surface quadrature,
\begin{equation}
 z_r=\sum_{j=1}^{N_s}A_j\,
 \psi_{\theta,r}(\bm{s}_j,\bm{n}_j,\omega)\,v_j,
 \label{eq:latent_source}
\end{equation}
and pressure is decoded as
\begin{equation}
 \widehat{p}_{\theta}(\bm{x}_q,\omega)
 =\frac{1}{\sqrt{R}}\sum_{r=1}^{R}
 \phi_{\theta,r}(\bm{x}_q,\omega)z_r.
 \label{eq:clbo_prediction}
\end{equation}
The factorized evaluation cost is $\mathcal{O}((N_s+N_r)R)$ after feature-network evaluation, rather than $\mathcal{O}(N_sN_r)$ for a dense source-receiver kernel. Figure~\ref{fig:clbo_architecture} summarizes the corresponding network structure.

\subsection{Exact structural properties}

\paragraph{Complex linearity.}
For fixed source coordinates, normals, quadrature weights, frequency, and receiver coordinates, $\bm{\psi}_{\theta}$ and $\bm{\phi}_{\theta}$ do not depend on source amplitude. Substitution of $av_1+bv_2$ into Eq.~\eqref{eq:latent_source} and distribution of the finite sum therefore yield
\begin{equation}
 \widehat{\Aop}_{\theta}(av_1+bv_2)
 =a\widehat{\Aop}_{\theta}(v_1)
 +b\widehat{\Aop}_{\theta}(v_2).
\end{equation}
The equality is exact in real arithmetic and is limited only by floating-point roundoff in implementation.

\paragraph{Zero-input consistency.}
Setting $v_j=0$ for all source nodes gives $z_r=0$ for every rank component and therefore $\widehat{p}_{\theta}=0$ without requiring training examples at zero excitation.

\paragraph{Permutation invariance.}
A common permutation of $\bm{s}_j$, $\bm{n}_j$, $A_j$, and $v_j$ leaves Eq.~\eqref{eq:latent_source} unchanged because the source aggregation is a sum.

\paragraph{Discretization interface.}
The input dimension is not tied to one number of source nodes. Changing the source discretization changes the finite quadrature in Eq.~\eqref{eq:latent_source}; accuracy then depends on both the learned kernel approximation and the quality of the source quadrature.

\subsection{Relation to prior operator structures}

The architecture shares ingredients with several established model classes. Because the vibrating boundary is represented by coordinates, normals, areas, and complex values, the source input is an unordered geometric point set rather than only a sampled vector. Its source summation has the permutation-invariant form studied in Deep Sets and is related to point-set and attention-based set encoders \cite{zaheer2017,qi2017pointnet,lee2019settransformer}; its source--receiver factorization is related to low-rank neural-operator kernels \cite{kovachki2023,koren2025svdno}; and its explicit integration weights are related to quadrature-based kernel operators \cite{lowery2024kno}. Geometry-informed operators additionally address variable domains and discretization convergence by encoding the domain itself \cite{li2023gino}, whereas the present model is conditioned on source and receiver coordinates inside one fixed cavity. Boundary-integral neural methods embed known integral equations or Green kernels into training \cite{fang2023boundary,qu2023}; in contrast, \clbo{} learns the acoustic transfer kernel from verified source--field pairs while hard-coding only the linear dependence on prescribed complex boundary velocity. Accordingly, the contribution is framed as a structure-preserving acoustic boundary-to-field formulation, supported by tests of exact source algebra and generalization to independently simulated complex source combinations.

\subsection{Fixed-sensor DeepONet baseline}

\begin{figure}[H]
\centering
\paperfigure{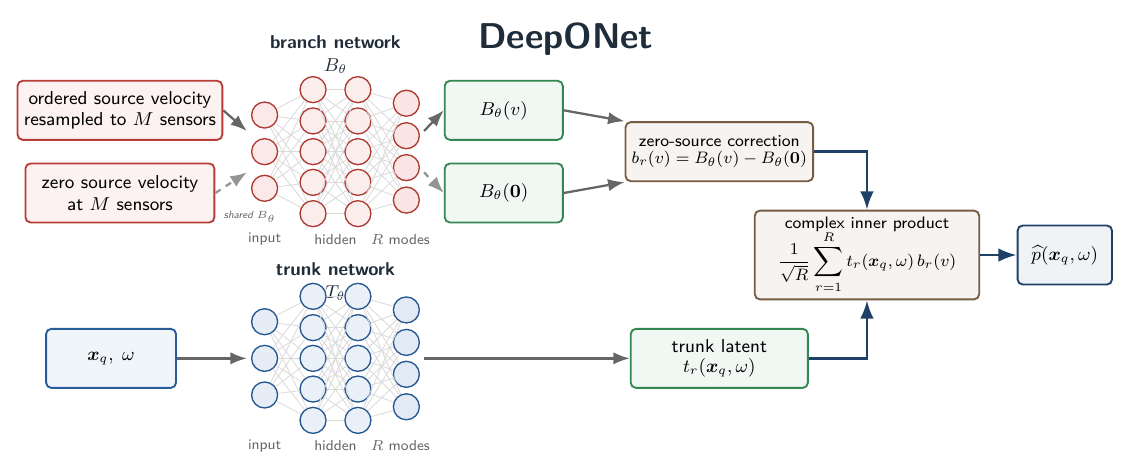}{Schematic of the fixed-sensor DeepONet baseline.}
\caption{Architecture of the fixed-sensor \deeponet{} baseline. The branch network encodes ordered real and imaginary source-velocity samples resampled to \BranchSensorCount{} fixed branch sensors when $N_s\neq M$. The same branch network is evaluated at zero input and subtracted so that zero wall velocity yields zero branch coefficients. The trunk network maps normalized receiver coordinates, frequency, and Fourier features to $R$ complex basis functions. Branch and trunk latents are combined by the same low-rank inner product as in Eq.~\eqref{eq:clbo_prediction}, but source amplitude remains inside the nonlinear branch encoder. The multilayer blocks are schematic; the implemented networks use depth~\NetworkDepth{}, hidden width~\HiddenWidth{}, and latent rank~\LatentRank{}.}
\label{fig:deeponet_architecture}
\end{figure}

The baseline follows the conventional branch--trunk construction \cite{lu2021}. Its branch network receives the ordered real and imaginary source-velocity samples at \BranchSensorCount{} fixed branch sensors. When the number of source samples differs, the stored one-dimensional ordering is linearly resampled to that branch-sensor count. The trunk network receives receiver coordinates and frequency, and a complex latent contraction produces pressure. A zero-source branch response is subtracted so that the baseline also returns zero for zero excitation. The branch remains nonlinear in source velocity and does not guarantee complex superposition or invariance to arbitrary source-node permutations. Both models use the same latent rank, hidden width, depth, coordinate features, optimizer schedule, data splits, and evaluation metrics. Because the input encoders differ, equal widths do not imply identical parameter counts; parameter counts and inference costs are therefore reported explicitly, and conclusions are based on both predictive error and structural tests rather than capacity alone. Figure~\ref{fig:deeponet_architecture} shows the baseline layout used in the comparison.

\section{Reference-data generation and verification}

\begin{figure}[H]
\centering
\paperfigure[0.82\textwidth]{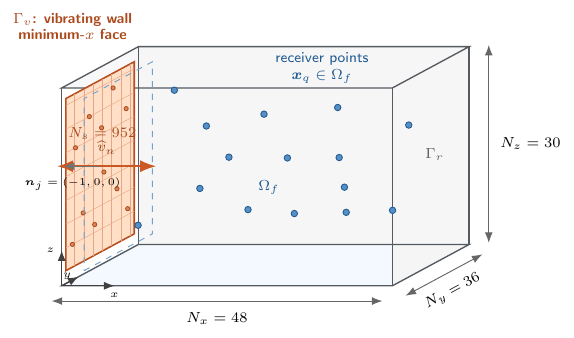}{Geometry and sampling layout of the rectangular reference cavity.}
\caption{Geometry and sampling layout used for reference-data generation. The rectangular \CavityShape{} lattice cavity has prescribed normal velocity on the minimum-$x$ wall $\Gamma_v$, rigid remaining walls $\Gamma_r$, $N_s=\SourceNodeCount{}$ source nodes on the vibrating wall, and interior receiver points $\bm{x}_q\in\Omega_f$.}
\label{fig:reference_geometry}
\end{figure}

Reference cases are generated in the fixed rectangular cavity shown in
Fig.~\ref{fig:reference_geometry}. The minimum-$x$ wall is prescribed as the
vibrating boundary, the remaining walls are rigid, and complex pressure is
sampled at interior receiver locations.

\subsection{MRT-LBM formulation}

Reference cases are generated with a three-dimensional 19-velocity (D3Q19) MRT-LBM. The collision step follows the standard MRT-LBM formulation \cite{lallemand2000,dhumieres2002}. LBM has also been benchmarked in broader direct-numerical-simulation/large-eddy-simulation (DNS/LES) flow settings \cite{yu2005dns,chikatamarla2010dns,khan2026boltzmann}, while acoustic-wave propagation and damping in LBM are addressed by acoustic-specific studies cited below. The collision step is performed in moment space,

\begin{align}
 \bm{m}(\bm{x},t) &= \bm{M}\bm{f}(\bm{x},t),\\
 \bm{m}^{*}(\bm{x},t) &=
 \bm{m}(\bm{x},t)
 -\bm{S}\left[\bm{m}(\bm{x},t)-\bm{m}^{\mathrm{eq}}(\bm{x},t)\right],\\
 \bm{f}^{*}(\bm{x},t) &= \bm{M}^{-1}\bm{m}^{*}(\bm{x},t),\\
 f_i(\bm{x}+\bm{c}_i\Delta t,t+\Delta t) &= f_i^{*}(\bm{x},t).
 \label{eq:mrt_lbm}
\end{align}
Here $\bm{M}$ is the moment transformation, $\bm{S}$ is diagonal, and conserved density and momentum moments are not relaxed. LBM acoustic-wave propagation and damping have been studied in prior acoustic LBM work \cite{viggen2011waves,astoul2021}. Low-dispersion and low-dissipation MRT-LBM formulations have also been developed specifically for computational aeroacoustics \cite{xu2011drplbm}. In the present reference solver, MRT collision is used to control numerical dispersion, dissipation, and stability.

The outer lattice shell represents the boundary. Five faces use halfway bounce-back and one face receives a harmonic moving-wall correction. A half-cosine ramp reduces broadband start-up content. Complex pressure is recovered by fitting the time history after a specified transient interval to the target harmonic. All quantities in the present dataset are reported in nondimensional lattice units; no physical sound-pressure level is inferred without a separate dimensional mapping.

\subsection{Verification protocol}

\begin{figure}[H]
\centering
\paperfigure{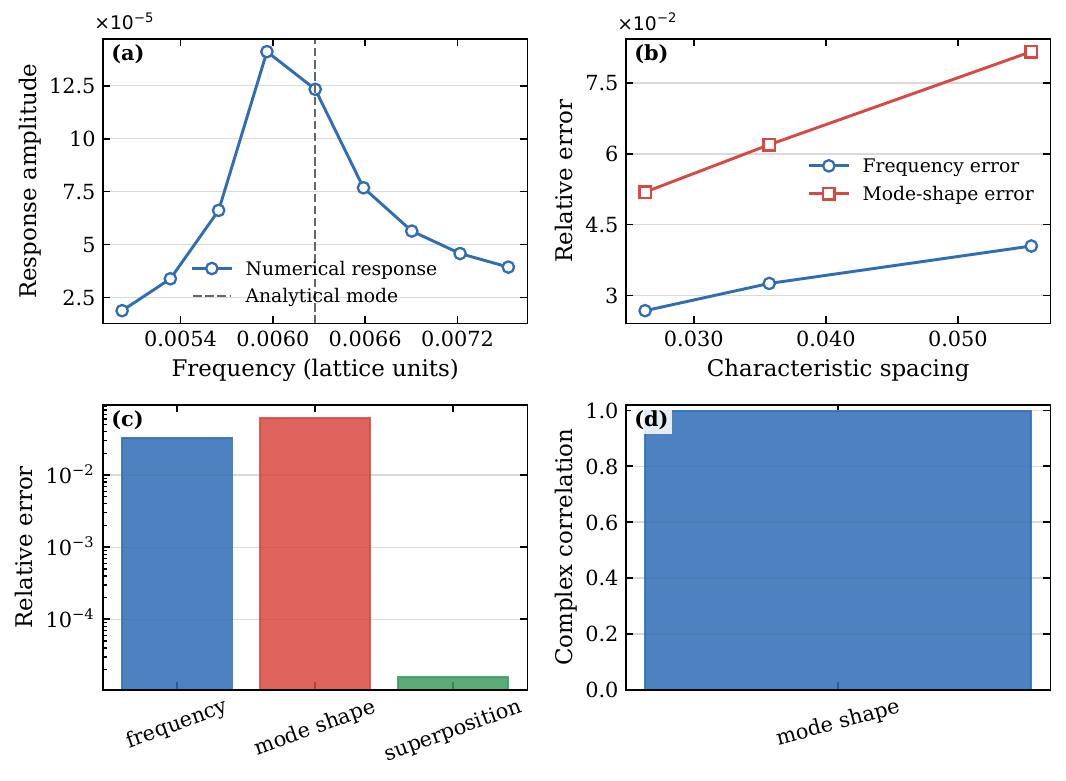}{Analytical and numerical resonance, mode-shape comparison, superposition check, and resolution study.}
\caption{Verification of the MRT-LBM reference solver. Frequencies and response amplitudes are reported in nondimensional lattice units; relative errors and correlations are dimensionless. Panel~(a) shows the forced response amplitude on the $48\times36\times30$ training cavity as the excitation frequency is swept about the first longitudinal mode; the dashed vertical line marks the analytical resonance frequency. Panel~(b) reports relative frequency error and mode-shape error versus normalized characteristic grid spacing, defined as the inverse of the shortest fluid-node count, for three rectangular resolutions ($32\times24\times20$, $48\times36\times30$, and $64\times48\times40$); the middle resolution corresponds to the training configuration. Panel~(c) lists the resonance frequency error, mode-shape error, and small-amplitude superposition error measured on the training grid. Panel~(d) gives the complex correlation between the numerical and analytical mode shapes after optimal complex scaling.}

\label{fig:solver_validation}
\end{figure}

The verification suite checks lattice identities and equilibrium preservation, a separate closed-duct piston benchmark that directly tests the moving-wall boundary condition without complex scaling, the first longitudinal rigid-cavity resonance on the $48\times36\times30$ training grid, the associated mode shape, small-amplitude source superposition, and a multi-resolution study on rectangular cavities with the same aspect ratio. For a rigid rectangular cavity with dimensions $L_x,L_y,L_z$, the analytical frequencies are
\begin{equation}
 f_{lmn}=\frac{c_0}{2}
 \sqrt{\left(\frac{l}{L_x}\right)^2+
       \left(\frac{m}{L_y}\right)^2+
       \left(\frac{n}{L_z}\right)^2}.
 \label{eq:cavity_modes}
\end{equation}
The corresponding pressure mode is proportional to
\begin{equation}
 \Phi_{lmn}(x,y,z)=
 \cos\!\left(\frac{l\pi x}{L_x}\right)
 \cos\!\left(\frac{m\pi y}{L_y}\right)
 \cos\!\left(\frac{n\pi z}{L_z}\right).
\end{equation}
Mode-shape comparisons allow one optimal complex scale because an eigenfunction is defined only up to complex amplitude and phase; they therefore assess cavity eigenstructure rather than the absolute amplitude and phase of the wall forcing, which is checked separately by the closed-duct piston benchmark. Figure~\ref{fig:solver_validation} summarizes the resonance check on the training grid and the normalized grid-resolution study across coarser and finer rectangular cavities.

Thus, the reference-data pipeline uses a two-stage validation hierarchy: analytical acoustic benchmarks verify the MRT-LBM solver for canonical cases, and the learned operators are then evaluated against this verified MRT-LBM reference over the full boundary-excitation dataset.

\subsection{Dataset configuration}

The principal dataset contains \DatasetCaseCount{} complete source-frequency cases. Each case stores source coordinates, normals, areas, complex velocity, frequency, receiver coordinates, complex pressure, \InteriorCollocationCount{} interior collocation points, and \RigidWallCollocationCount{} rigid-wall collocation points. Table~\ref{tab:dataset_configuration} summarizes the resolved generation configuration.

\begin{table}[H]
\centering
\caption{Resolved reference-data configuration. Frequencies and field quantities are expressed in lattice units.}
\label{tab:dataset_configuration}
\begin{tabular}{p{5.6cm}p{6.8cm}}
\toprule
Quantity & Value \\
\midrule
Geometry & \CavityGeometry{} cavity \\
Lattice shape & \CavityShape \\
Vibrating boundary & minimum-$x$ wall \\
Source nodes on reference wall & \SourceNodeCount \\
MRT relaxation parameter $\tau$ & \LBMTau \\
Time steps per case & \LBMSteps \\
Harmonic-fit start step & \LBMFitStart \\
Source-ramp duration & \LBMRampSteps \\
Cases & \DatasetCaseCount \\
Discrete frequencies & \FrequencyCount \\
Frequency range & \FrequencyRange \\
Receiver queries per case & \ReceiversPerCase \\
Source families & Gaussian, Fourier, smooth random modal; uniform wall motion used only for piston verification \\
Peak source-velocity range & $2\times10^{-5}$--$8\times10^{-5}$ lattice velocity units \\
Stored pressure & complex lattice pressure \\
\bottomrule
\end{tabular}
\end{table}

\section{Learning and evaluation protocol}

\subsection{Case-level splitting and normalization}

Complete source-frequency cases are split before model training. The default split uses \TrainingFraction{} of cases for training, \ValidationFraction{} for validation, and \TestingFraction{} for testing. All receiver samples from one case remain in the same split. A single fixed case-level split is reused across models for the principal comparison. Coordinate, frequency, source-velocity, and pressure statistics are computed from the training split only. Each case stores \ReceiversPerCase{} receiver queries. The configured training cap is \MaxReceiverPoints{} receiver points, so all stored receivers are used during both training and evaluation.

The main comparison is repeated for \RandomSeedCount{} seeds (\(2026, 1234, 2345, 3456, 4567\)). Results are reported as mean and standard deviation across seeds, with paired per-case distributions retained for statistical comparison. The source-family holdout is a separate seed~2026 secondary study in which the smooth random modal source family is excluded from training and validation and used only for testing.

\subsection{Training objective}

The primary loss is a masked complex mean-square error over receiver queries,
\begin{equation}
 \mathcal{L}_{\mathrm{data}}
 =\frac{1}{2N}\sum_{q=1}^{N}
 \left\|\widehat{\bm{p}}_{\theta,q}-\bm{p}_{q}\right\|_2^2,
 \label{eq:data_loss}
\end{equation}
where $\bm{p}=(\Re p,\Im p)$. The principal model comparison uses supervised training only (\(\lambda_H=\lambda_R=\lambda_V=0\)) so that the architectural effect is not confounded by derivative-based penalties.

Physics-informed variants are evaluated as ablations. The training implementation supports normalized residuals for the Helmholtz equation, the rigid-wall Neumann condition, and the prescribed vibrating-wall condition,
\begin{equation}
 \mathcal{L}=\lambda_d\mathcal{L}_{\mathrm{data}}
 +\lambda_H\mathcal{L}_{H}
 +\lambda_R\mathcal{L}_{R}
 +\lambda_V\mathcal{L}_{V}.
 \label{eq:combined_loss}
\end{equation}
The principal model comparison uses \(\lambda_H=\lambda_R=\lambda_V=0\). The physics-residual ablation reported in Table~\ref{tab:ablation_results} and Fig.~\ref{fig:ablation} activates only the interior Helmholtz and rigid-wall terms (\(\lambda_H=2.25\times10^{-3}\), \(\lambda_R=10^{-3}\), \(\lambda_V=0\)) in the physics-residual training configuration. During preliminary weight calibration, adding \(\mathcal{L}_{V}\) consistently degraded validation performance relative to the Helmholtz--rigid configuration, so the vibrating-wall residual was omitted from the reported analysis. The residual weights and ramp schedules are held fixed within each ablation.

\subsection{Model and optimization settings}

\begin{table}[H]
\centering
\caption{Common model and optimization settings for the principal comparison.}
\label{tab:training_configuration}
\begin{tabular}{p{5.8cm}p{6.6cm}}
\toprule
Setting & Value \\
\midrule
Latent rank & \LatentRank \\
Hidden width & \HiddenWidth \\
Hidden layers per network & \NetworkDepth \\
Activation & hyperbolic tangent \\
Coordinate Fourier bands & \CoordinateFourierBands \\
Frequency Fourier bands & \FrequencyFourierBands \\
DeepONet branch sensors & \BranchSensorCount \\
Training receiver cap & \MaxReceiverPoints{} points \\
Cases per batch & \TrainingBatchCases \\
Maximum epochs & \TrainingEpochs \\
Optimizer & AdamW \\
Base learning rate & \InitialLearningRate \\
Minimum learning rate & $2\times10^{-6}$ \\
Weight decay & \WeightDecay \\
Warm-up & 10 epochs \\
Gradient clipping & 1.0 \\
Early-stopping patience & 60 epochs \\
Training precision & single precision (\texttt{float32}) \\
\bottomrule
\end{tabular}
\end{table}

\subsection{Evaluation studies}

The evaluation is organized around distinct questions rather than one aggregate random-split score.

\begin{enumerate}[leftmargin=1.8em]
\item \textbf{Primary held-out cases:} accuracy on test source-frequency cases using all stored receiver queries.
\item \textbf{Receiver interpolation:} both models are retrained using only a deterministic subset of receiver coordinates in each case, and errors are evaluated only on the held-out receiver coordinates.
\item \textbf{Algebraic structure:} complex superposition, homogeneity, zero-input response, and source-node permutation.
\item \textbf{Unseen source combinations:} held-out source fields at a common frequency are combined using random complex coefficients; for each combined boundary excitation, a fresh MRT-LBM reference solution is generated and evaluated on a common receiver cloud.
\item \textbf{Source discretization:} the same source-generation rule sampled on the native, coarsened, and nonuniform source-wall node sets while the physical surface is held fixed.
\item \textbf{Source-family generalization:} the smooth random modal source family is excluded from training and validation and used only for testing.
\item \textbf{Data efficiency and physics ablations:} training with 10\%, 25\%, 50\%, and 100\% of the available training cases, with optional residual penalties.
\item \textbf{Computational performance:} synchronized inference timing at fixed source and receiver counts, together with reference-solver time, parameter count, and peak memory.
\end{enumerate}

\subsection{Metrics}

The primary metric is the complex relative field error
\begin{equation}
 \epsilon_{p}=\frac{\|\widehat{p}-p\|_2}{\|p\|_2},
 \label{eq:relative_error}
\end{equation}
evaluated over valid receiver samples in a case or over a pooled evaluation vector formed by concatenating all valid receiver points in a split. Secondary reported metrics are complex correlation,
\begin{equation}
 \rho_c =
 \frac{\left|\sum_q p_q^{*}\widehat{p}_q\right|}
 {\left(\sum_q |p_q|^2\right)^{1/2}
  \left(\sum_q |\widehat{p}_q|^2\right)^{1/2}},
 \label{eq:complex_correlation}
\end{equation}
amplitude-ratio mean absolute error in decibels,
\begin{equation}
 \epsilon_{A,\mathrm{dB}}=
 \frac{1}{N_m}\sum_{q\in\mathcal{M}}
 \left|20\log_{10}\frac{|\widehat{p}_q|+\varepsilon}{|p_q|+\varepsilon}\right|,
\end{equation}
and phase mean absolute error
\begin{equation}
 \epsilon_{\varphi}=
 \frac{1}{N_m}\sum_{q\in\mathcal{M}}
 \left|\arg\!\left(\widehat{p}_q p_q^{*}\right)\right|.
\end{equation}
Phase errors are reported in degrees in the result tables and figures. The mask $\mathcal{M}$ contains valid receiver samples whose reference amplitude is within 60~dB of the maximum reference amplitude of the evaluated field or pooled evaluation vector. Table~\ref{tab:main_results} reports pooled split values for $\epsilon_p$, $\epsilon_{A,\mathrm{dB}}$, and $\epsilon_{\varphi}$, and the mean of per-case values for $\rho_c$. Algebraic superposition is measured by
\begin{equation}
 \epsilon_{\mathrm{lin}}=
 \frac{\|\widehat{\Aop}(av_1+bv_2)
 -a\widehat{\Aop}(v_1)-b\widehat{\Aop}(v_2)\|_2}
 {\|a\widehat{\Aop}(v_1)+b\widehat{\Aop}(v_2)\|_2+\varepsilon}.
 \label{eq:linearity_metric}
\end{equation}

The algebraic test above verifies implementation structure but does not by itself establish predictive accuracy on a new physical field. A separate mixed-source test constructs
\begin{equation}
 v_{\mathrm{mix}}=a v_1+b v_2,
 \label{eq:mixed_source}
\end{equation}
from held-out source fields at the same frequency, carries out a new MRT-LBM reference simulation, and evaluates
\begin{equation}
 \epsilon_{\mathrm{mix}}=
 \frac{\|\widehat{p}(v_{\mathrm{mix}})-p_{\mathrm{LBM}}(v_{\mathrm{mix}})\|_2}
 {\|p_{\mathrm{LBM}}(v_{\mathrm{mix}})\|_2}.
 \label{eq:mixed_source_error}
\end{equation}
The paired difference between the fixed-sensor \deeponet{} baseline and \clbo{} errors is summarized with a 95\% percentile bootstrap interval over the identical mixed cases, using 10,000 resamples of the paired case-level differences \cite{efron1994}.

\section{Results}

\subsection{Reference-solver verification}

Figure~\ref{fig:solver_validation} reports the checks that were completed before the $640$-case generator was enabled. In panel~(a), the numerical response peaks near $f=0.0061$ while the analytical first longitudinal mode lies at $f=0.0063$; the resulting relative frequency error is \ModeFrequencyError{}. Panel~(b) shows that both frequency and mode-shape errors decrease as the rectangular cavity resolution is refined from $32\times24\times20$ to $64\times48\times40$ nodes, with the $48\times36\times30$ training resolution in between. The horizontal coordinate is a normalized characteristic grid spacing, $1/\min(N_x-2,N_y-2,N_z-2)$, rather than a change to the unit LBM lattice spacing; fitting frequency error against this spacing gives an observed order of \ConvergenceOrder{}.

Panel~(c) summarizes the training-grid verification metrics: mode-shape relative error \ModeShapeError{} and superposition error \SolverLinearityError{}. Panel~(d) confirms strong spatial agreement between numerical and analytical mode shapes, with complex correlation \ModeShapeCorrelation{}.

A separate closed-duct piston validation directly checks the prescribed normal-velocity sign and phasor convention without fitting a complex scale. At $f/f_1=\PistonFrequencyRatio{}$, the complex relative pressure error was \PistonComplexError{}, the amplitude relative error was \PistonAmplitudeError{}, and the phase root-mean-square error (RMSE) was \PistonPhaseRMSE{}°. Because the analytical duct solution is lossless whereas the MRT-LBM calculation at $\tau=\LBMTau{}$ has finite viscosity and finite-grid dissipation, exact inviscid amplitude agreement is not expected; the result is therefore interpreted as a finite-grid sign, phase, and amplitude-consistency check.

This piston benchmark corresponds to the uniform normal-velocity limit of the vibrating wall; nonuniform Gaussian, Fourier, and smooth random modal source fields are assessed through the verified MRT-LBM reference data rather than by comparison with the one-dimensional piston solution.

\subsection{Primary predictive accuracy}

\begin{figure}[H]
\centering
\paperfigure{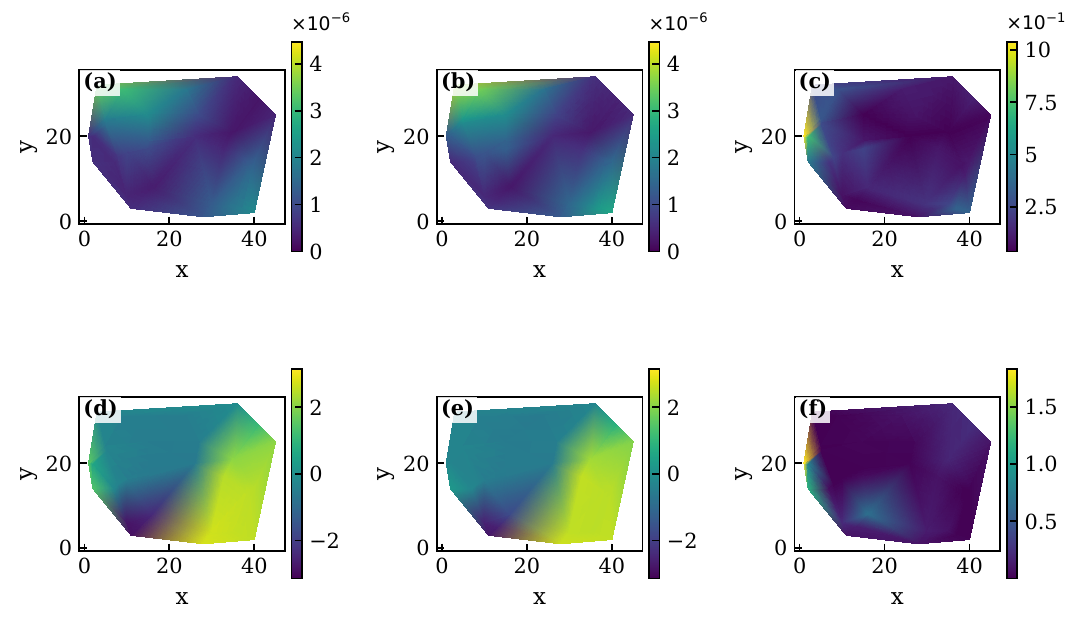}{Held-out test case: reference and predicted magnitude, normalized error, and phase on a receiver-plane slice.}
\caption{Complex pressure predictions for one held-out test case near the median per-case test error. Fields are shown on the fixed-$z$ receiver plane with the densest sampling and rendered by triangulation over scattered receiver locations; they are not a structured interior grid. Panels~(a) and~(b) compare MRT-LBM reference and \clbo{} predicted magnitude on a shared color scale. Panel~(c) shows the pointwise complex error magnitude normalized by the case root-mean-square reference amplitude. Panels~(d)--(f) show reference phase, predicted phase, and absolute wrapped phase error in radians, respectively, masked where the amplitude mask $\mathcal{M}$ applies.}
\label{fig:prediction_fields}
\end{figure}

\begin{figure}[H]
\centering
\paperfigure{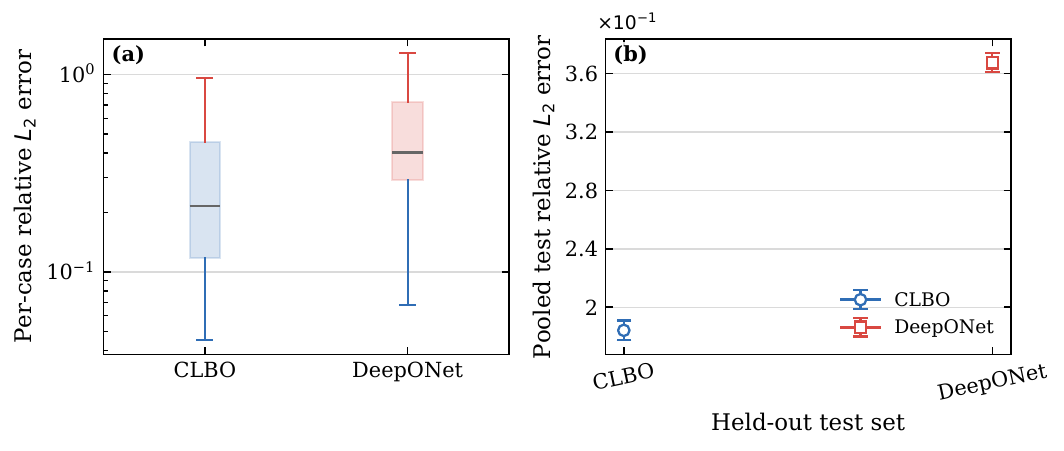}{Per-case and pooled held-out test errors for \clbo{} and the fixed-sensor \deeponet{}.}
\caption{Primary model comparison on the shared held-out test split. Both models were trained on the \DatasetCaseCount{}-case \CavityShape{} dataset with identical source fields, receiver clouds, optimization settings, and supervised-only losses. Panel~(a) shows the per-case complex relative field error for all \TestingFraction{} test cases, pooled over \RandomSeedCount{} training seeds; each box summarizes the distribution over held-out cases and training seeds. Panel~(b) shows the global test-set relative error obtained by concatenating all receiver points in the test split and averaging over the same seeds; it uses the same pooled relative-$L_2$ definition as the validation metric used for early stopping and is the test metric reported in Table~\ref{tab:main_results}. Markers give the seed-averaged pooled error and bars indicate 95\% confidence intervals across seeds.}
\label{fig:model_comparison}
\end{figure}

Across \RandomSeedCount{} seeds, \clbo{} achieved a complex relative field error of \CLBORelativeError{} $\pm$ \CLBORelativeErrorStd{}, while the fixed-sensor \deeponet{} obtained \DeepONetRelativeError{} $\pm$ \DeepONetRelativeErrorStd{}. The corresponding amplitude-ratio errors were \CLBOAmplitudeError{} and \DeepONetAmplitudeError{}, and the phase errors were \CLBOPhaseError{} and \DeepONetPhaseError{}. Table~\ref{tab:main_results} reports the full comparison. Figure~\ref{fig:prediction_fields} visualizes one held-out test case chosen near the median per-case error on the test split. The MRT-LBM reference and the supervised \clbo{} prediction are evaluated at the same receiver coordinates and displayed on the fixed-$z$ plane with the densest receiver sampling. Panel~(a) shows the reference magnitude and panel~(b) the predicted magnitude; the two panels share a common color scale. Panel~(c) maps the pointwise complex error normalized by the case root-mean-square reference amplitude. Panels~(d) and~(e) show reference and predicted phase, and panel~(f) the absolute wrapped phase error, masked where the amplitude mask $\mathcal{M}$ applies. Figure~\ref{fig:model_comparison} summarizes the same held-out evaluation in aggregate. Panel~(a) pools per-case errors over \RandomSeedCount{} training seeds and therefore sits slightly above the pooled values when hard resonant cases are present. Panel~(b) reports the global test-set error obtained by concatenating all receiver points in the split; the markers match Table~\ref{tab:main_results} and the early-stopping criterion.

\begin{table}[H]
\centering
\caption{Primary held-out-case performance. Relative $L_2$ values are mean $\pm$ standard deviation across independent training seeds; other reported metrics are seed means. Amplitude error is an amplitude-ratio error and is not a physical sound pressure level (SPL).}
\label{tab:main_results}
\small
\resizebox{\textwidth}{!}{%
\begin{tabular}{lccccc}
\toprule
Model & Relative $L_2$ & Complex corr. & Amp. MAE (dB) & Phase MAE (deg) & Parameters \\
\midrule
Fixed-sensor \deeponet{} &
\DeepONetRelativeError{} $\pm$ \DeepONetRelativeErrorStd{} &
\DeepONetComplexCorrelation{} &
\DeepONetAmplitudeError{} &
\DeepONetPhaseError{} &
\DeepONetParameterCount{} \\
\clbo{} &
\CLBORelativeError{} $\pm$ \CLBORelativeErrorStd{} &
\CLBOComplexCorrelation{} &
\CLBOAmplitudeError{} &
\CLBOPhaseError{} &
\CLBOParameterCount{} \\
\bottomrule
\end{tabular}%
}
\end{table}

On five newly simulated mixed-source cases, evaluated with the seed~2026 primary supervised checkpoints, the mean relative errors were \CLBOUnseenCombinationError{} for \clbo{} and \DeepONetUnseenCombinationError{} for \deeponet{}. \clbo{} achieved a paired win rate of \CLBOUnseenWinRate{}. The mean paired advantage, defined as $\epsilon_{\mathrm{DeepONet}}-\epsilon_{\mathrm{CLBO}}$, was \UnseenPairedAdvantage{} with a 95\% percentile bootstrap interval of [\UnseenAdvantageCILower{}, \UnseenAdvantageCIUpper{}] over the five paired mixed cases. Because each target was generated by a new MRT-LBM reference simulation, this test measures predictive generalization to unseen complex source mixtures rather than only algebraic consistency of model outputs.

\begin{table}[H]
\centering
\caption{Generalization to newly simulated complex source mixtures. Pairs are drawn from the held-out test split at common frequency and source mesh; each mixture is simulated once with MRT-LBM and evaluated on the same receiver cloud. Both models use the seed~2026 primary supervised checkpoints. The manuscript reports the paired error advantage with a 95\% percentile bootstrap interval over paired case-level differences; win rate counts mixtures with lower relative $L_2$ error.}
\label{tab:unseen_results}
\small
\resizebox{\textwidth}{!}{%
\begin{tabular}{lccccc}
\toprule
Model & Mixed cases & Mean relative $L_2$ & Median relative $L_2$ & Paired win rate & Mean complex corr. \\
\midrule
Fixed-sensor \deeponet{} & \UnseenCombinationCount{} & \DeepONetUnseenCombinationError{} & \DeepONetUnseenMedianError{} & \DeepONetUnseenWinRate{} & \DeepONetUnseenComplexCorrelation{} \\
\clbo{} & \UnseenCombinationCount{} & \CLBOUnseenCombinationError{} & \CLBOUnseenMedianError{} & \CLBOUnseenWinRate{} & \CLBOUnseenComplexCorrelation{} \\
\bottomrule
\end{tabular}%
}
\end{table}

\subsection{Receiver interpolation}

Receiver-coordinate interpolation was evaluated by retraining both models with only a deterministic subset of receiver locations included in the supervised loss. The remaining receiver locations were excluded during training and used only for evaluation. This test measures whether the learned surrogate represents a continuous pressure field over receiver coordinates rather than only fitting the receiver locations used in training.

\begin{figure}[H]
\centering
\paperfigure[0.58\textwidth]{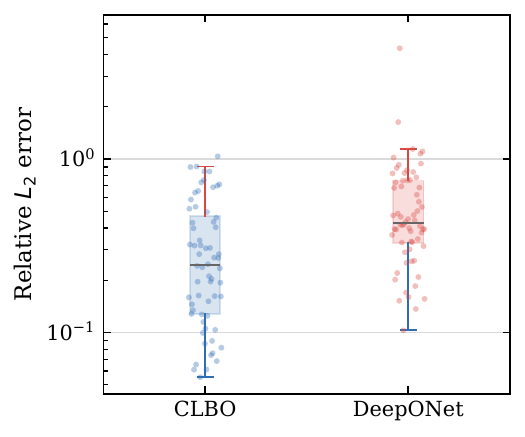}{Receiver-interpolation errors on held-out receiver coordinates.}
\caption{Receiver interpolation on held-out receiver coordinates. Both models were retrained using only a deterministic subset of receiver locations in each case and evaluated on the remaining unseen receiver locations inside the same cavity. Each point is one held-out test case ($n=\ReceiverInterpolationCaseCount{}$). \clbo{} gives lower mean and median relative field error than the fixed-sensor \deeponet{} baseline.}
\label{fig:receiver_interpolation}
\end{figure}

Across \ReceiverInterpolationCaseCount{} held-out test cases, \clbo{} reduced the mean held-out-receiver relative error from \DeepONetReceiverInterpolationMean{} for \deeponet{} to \CLBOReceiverInterpolationMean{}. The median error was also reduced, from \DeepONetReceiverInterpolationMedian{} for \deeponet{} to \CLBOReceiverInterpolationMedian{} for \clbo{}. Figure~\ref{fig:receiver_interpolation} shows the per-case error distributions on the held-out receiver coordinates.

\subsection{Exact structure and source discretization}

\begin{figure}[H]
\centering
\paperfigure{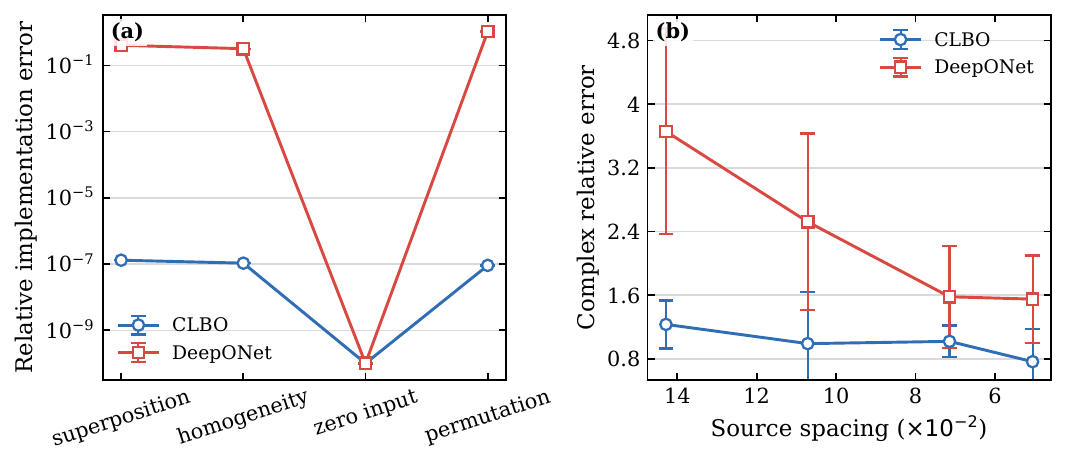}{Algebraic structure tests and fixed-geometry source-remeshing accuracy.}
\caption{Algebraic structure and source-wall sampling transfer on the primary supervised checkpoints. Panel~(a) shows superposition, homogeneity, zero-input, and permutation implementation errors measured with the same inference interface used in training; exact zero-input responses are shown at the logarithmic floor ($10^{-10}$). Panel~(b) shows complex relative error on alternative source-wall node subsets at fixed geometry, excluding the native training discretization; characteristic source-wall sampling spacing decreases from left to right. Markers in panel~(b) give means over remeshed test cases and bars indicate 95\% confidence intervals across cases.}

\label{fig:structure_remeshing}
\end{figure}

The measured \clbo{} superposition error was \CLBOLinearityError{}, compared with \DeepONetLinearityError{} for \deeponet{}. The zero-input and source-permutation errors of \clbo{} were \CLBOZeroInputError{} and \CLBOPermutationError{}, respectively. These measurements confirm the algebraic guarantees of Eqs.~\eqref{eq:latent_source}--\eqref{eq:clbo_prediction} at implementation precision. Figure~\ref{fig:structure_remeshing} reports the same tests in panel~(a).

When the same source-generation rule was sampled on alternative source-wall node subsets, \clbo{} obtained a source-sampling transfer error of \CLBOMeshTransferError{}, whereas the fixed-sensor baseline obtained \DeepONetMeshTransferError{}. The experiment used the same physical domain and receiver cloud while changing the source-wall sampling pattern. Panel~(b) of Figure~\ref{fig:structure_remeshing} summarizes the spacing-resolved remeshing errors.

\subsection{Source-family holdout}

The source-family holdout evaluates generalization to a source-field family not seen during training. In this secondary seed~2026 study, the smooth random modal source family is removed from both training and validation. The resulting training and validation splits contain only Gaussian and Fourier source fields, while the held-out test split contains 18 smooth random modal cases.

The mean per-case complex relative errors were \CLBOSourceFamilyError{} for \clbo{} and \DeepONetSourceFamilyError{} for \deeponet{}. This result indicates that \clbo{} gives lower error than the fixed-sensor \deeponet{} baseline when evaluated on the unseen smooth random modal source family.

\subsection{Ablations and data efficiency}

\begin{figure}[H]
\centering
\paperfigure{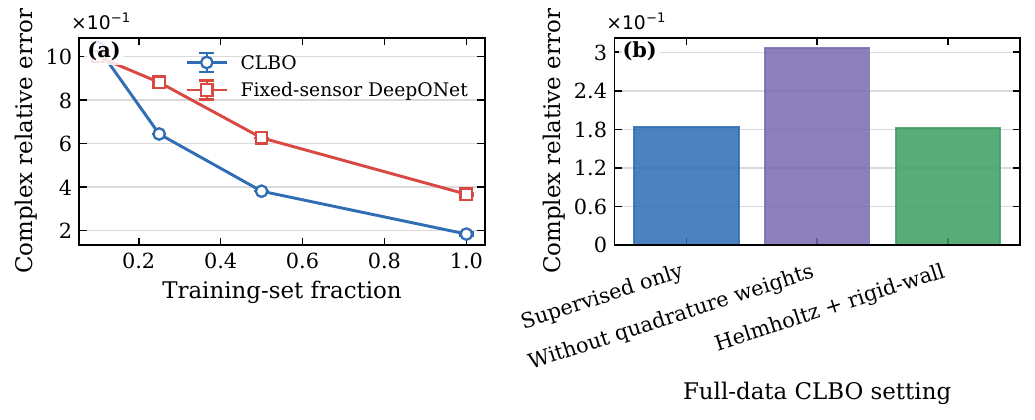}{Training-set fraction curves and full-data CLBO ablations.}
\caption{Data efficiency and controlled ablations on the fixed test manifest. Panel~(a) shows complex relative error versus training-set fraction for \clbo{} and the fixed-sensor \deeponet{}; points at 10\%, 25\%, and 50\% are single-seed (2026) data-efficiency training evaluations, and the 100\% endpoints report five-seed primary test means. Panel~(b) compares full-data \clbo{} settings: the five-seed supervised-only primary mean, an inference-time source-area ablation that sets quadrature weights to one using the seed~2026 supervised checkpoint, and a seed~2026 model trained with Helmholtz and rigid-wall physics residuals \((\lambda_V=0)\).}

\label{fig:ablation}
\end{figure}

With 10\% of the training cases, \clbo{} and \deeponet{} obtained errors of \CLBOTenPercentError{} and \DeepONetTenPercentError{}, respectively. Table~\ref{tab:ablation_results} reports the full training-fraction grid. Figure~\ref{fig:ablation} summarizes the same evaluation graphically. Panel~(a) shows how test error decreases with training-set fraction for \clbo{} and the fixed-sensor \deeponet{}. Panel~(b) compares three full-data \clbo{} settings: supervised-only training, an inference-time ablation that removes quadrature weights from the source contraction (\NoQuadratureError{}), and training with Helmholtz and rigid-wall physics residuals (\PhysicsCLBOError{}). The vibrating-wall residual was excluded from the reported physics-residual training configuration after preliminary calibration showed it degraded validation performance. These ablations separate the effect of architectural linearity, numerical quadrature, and derivative-based regularization.

\begin{table}[H]
\centering
\caption{Ablation and data-efficiency results on the fixed test manifest. Entries are complex relative field errors. Columns labeled 10\%, 25\%, and 50\% denote the fraction of the 512-case training split used during training: whole training cases were drawn once at random (seed~2026) while validation and test splits stayed fixed. Supervised \clbo{} and \deeponet{} entries at 100\% report the five-seed primary test means from Table~\ref{tab:main_results}. The quadrature-removal and physics-residual \clbo{} variants were evaluated at full training data only (seed~2026); the physics-residual row uses Helmholtz and rigid-wall penalties only (\(\lambda_V=0\)).}
\label{tab:ablation_results}
\small
\resizebox{\textwidth}{!}{%
\begin{tabular}{lcccc}
\toprule
Variant & 10\% of train & 25\% of train & 50\% of train & 100\% of train \\
\midrule
Fixed-sensor \deeponet{} & \DeepONetTenPercentError{} & \DeepONetTwentyFivePercentError{} & \DeepONetFiftyPercentError{} & \DeepONetRelativeError{} \\
\clbo{} without areas & --- & --- & --- & \NoQuadratureError{} \\
\clbo{} & \CLBOTenPercentError{} & \CLBOTwentyFivePercentError{} & \CLBOFiftyPercentError{} & \CLBORelativeError{} \\
\clbo{} with Helmholtz and rigid-wall residuals & --- & --- & --- & \PhysicsCLBOError{} \\
\bottomrule
\end{tabular}%
}
\end{table}

\subsection{Computational performance}

\begin{figure}[H]
\centering
\paperfigure{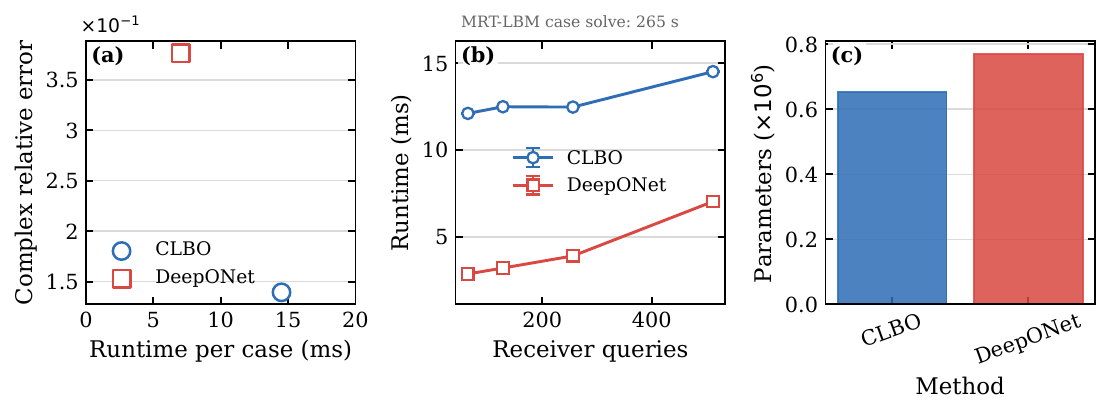}{Accuracy-runtime tradeoff, receiver-query scaling, and parameter counts.}
\caption{Accuracy and computational cost from the inference benchmark on the primary supervised checkpoints. Panel~(a) compares complex relative error against per-case neural inference time at the full receiver count ($512$ queries, $952$ source nodes). Panel~(b) shows how neural runtime scales with receiver-query count; the MRT-LBM reference solve time is a single full lattice simulation and is annotated because it does not scale with query count. Panel~(c) reports trainable parameter counts for the supervised checkpoints; peak device memory was not measured because the benchmark was run on central processing unit (CPU).}
\label{fig:runtime}
\end{figure}

On \HardwareDescription{}, with $952$ source nodes and $512$ receiver queries on one held-out test case, \clbo{} required \CLBOInferenceTime{}~ms per case and the fixed-sensor \deeponet{} required \DeepONetInferenceTime{}~ms. A separate MRT-LBM reference benchmark, re-running the same case with the data-generation solver, required \LBMCaseTime{}~ms. The resulting speedups relative to MRT-LBM were \CLBOSpeedup{} for \clbo{} and \DeepONetSpeedup{} for \deeponet{}. Neural timings exclude file input/output, use 10 warm-up forward passes followed by 100 timed forward passes, and apply device synchronization when applicable; the reference timing uses one untimed warm-up solve followed by two timed full lattice simulations. Table~\ref{tab:runtime_results} lists the measured per-case times and speedups. Figure~\ref{fig:runtime} summarizes the same benchmark graphically: panel~(a) compares accuracy against per-case neural inference time at the full receiver count, panel~(b) shows the scaling of neural runtime with receiver-query count while the MRT-LBM reference remains a fixed full-case solve, and panel~(c) compares the trainable parameter counts of the supervised checkpoints.

\begin{table}[H]
\centering
\caption{Computational cost at $952$ source nodes and $512$ receiver queries on one held-out test case, using the seed~2026 primary supervised checkpoints. Neural timings exclude file input/output and report the mean of 100 timed forward passes after 10 warm-up passes, with device synchronization applied when applicable. The MRT-LBM reference time is a separate benchmark using one untimed warm-up solve followed by two timed full lattice solves on the same case. Peak device memory was not recorded on CPU.}

\label{tab:runtime_results}
\small
\resizebox{\textwidth}{!}{%
\begin{tabular}{lccc}
\toprule
Method & Time per case & Peak memory & Speed relative to MRT-LBM \\
\midrule
MRT-LBM reference & \LBMCaseTime{} ms & N/A (CPU) & $1\times$ \\
Fixed-sensor \deeponet{} & \DeepONetInferenceTime{} ms & N/A (CPU) & \DeepONetSpeedup{} \\
\clbo{} & \CLBOInferenceTime{} ms & N/A (CPU) & \CLBOSpeedup{} \\
\bottomrule
\end{tabular}%
}
\end{table}

\section{Discussion}

The central distinction between \clbo{} and the fixed-sensor baseline is structural rather than merely architectural capacity. This emphasis is consistent with the broader move toward operator models that encode discretization, geometry, or integral structure rather than treating every physical relation as an unconstrained regression problem \cite{kovachki2023,li2023gino,lowery2024kno}. A generic nonlinear branch network must infer the superposition law from sampled source fields. In \clbo, that law is satisfied independently of training-set coverage because source amplitude appears only in the linear quadrature contraction. The measured error in Eq.~\eqref{eq:linearity_metric} therefore tests numerical implementation rather than learned behavior. The mixed-source experiment in Eq.~\eqref{eq:mixed_source_error} is the complementary predictive test: it determines whether the hard constraint improves accuracy on new source amplitudes and phases when the target field is generated independently.

Quadrature serves a separate purpose. Exact linearity alone does not guarantee invariance to a changed source mesh; the discrete sum must also approximate the same continuous boundary integral. Coordinates and areas provide the required interface, while the remeshing study measures the remaining learned-kernel and quadrature errors. The result should be interpreted as transfer across the tested discretizations, not as universal mesh independence.

The fixed-sensor \deeponet{} is an appropriate reference for the cost of replacing explicit source integration with a nonlinear function encoder. It is deliberately tied to an ordered sensor vector and therefore represents the common deployment condition in which the source mesh is fixed. The comparison isolates whether exact source structure is beneficial when both models use the same training cases, receiver trunk representation, latent rank, and optimization budget. These results should be interpreted within the scoped boundary-excitation setting: \clbo{} gains its advantage by embedding the complex-linear source algebra of the acoustic operator, whereas the fixed-sensor \deeponet{} baseline must approximate that structure from data.

Physics-informed residuals are secondary to the architectural contribution. They can regularize low-data training, but they add first- and second-order automatic-differentiation cost and introduce loss-balancing choices. In the present study, only the interior Helmholtz and rigid-wall residuals were retained in the reported physics ablation; the prescribed vibrating-wall term degraded validation performance during preliminary calibration and was set to zero. The supervised \clbo{} is therefore the principal model, and residual terms are retained only when the ablation demonstrates a consistent improvement.

The present limitations are explicit. The acoustic problem is linear and time harmonic, the medium is quiescent, and one trained model represents one fixed cavity and passive boundary configuration. The reference labels use nondimensional lattice units. The workflow itself is not tied to MRT-LBM: the stored data contract contains only surface geometry, complex boundary values, operating frequency, receiver coordinates, and complex target fields. Replacing the label generator does not change the model interface, but new geometries or boundary conditions require representative training data or an explicitly geometry-conditioned extension.

A natural extension is to couple the present complex-linear boundary operator with a geometry encoder, such as a point-cloud, graph, signed-distance, mesh-based, or multiscale geometric representation. Mesh- and graph-based learned simulators provide one possible route for encoding irregular geometric discretizations while retaining physical simulation structure \cite{pfaff2021meshgraphnets}. Such a model would condition the learned source and receiver bases on the passive acoustic domain while preserving the exact linear dependence on prescribed boundary velocity. This would extend the present fixed-cavity formulation toward geometry-varying acoustic surrogate modeling for industrial computer-aided engineering (CAE) workflows.

\section{Conclusions}

A quadrature-aware complex-linear neural operator was developed for distributed boundary-to-field prediction in resonant acoustics. The model combines learned source and receiver bases with an explicit complex surface contraction. This construction guarantees source superposition, homogeneity, zero-input consistency, and source-node permutation invariance while permitting variable source-sample counts.

Using verified MRT-LBM reference cases, \clbo{} achieved a mean complex relative error of \CLBORelativeError{} $\pm$ \CLBORelativeErrorStd{} across \RandomSeedCount{} seeds, compared with \DeepONetRelativeError{} $\pm$ \DeepONetRelativeErrorStd{} for the fixed-sensor \deeponet{}. Its superposition error was \CLBOLinearityError{}. On newly simulated mixed-source cases, the corresponding errors were \CLBOUnseenCombinationError{} and \DeepONetUnseenCombinationError{}. Its measured inference speedup was \CLBOSpeedup{}. The structured evaluations separate ordinary predictive accuracy from exact algebraic behavior, unseen source-combination generalization, source discretization, source-family holdout, label efficiency, and computational cost. The resulting workflow provides a structured boundary-to-field learning formulation whose data interface is independent of the method used to obtain verified reference fields.

\section*{Data availability}
The data supporting the findings of this study are available from the corresponding author upon reasonable request.

\section*{Acknowledgements}
This work was supported by Vinnova--Fordonsstrategisk forskning och innovation (FFI) in the project "3D Virtual Platform for Digitalization of Holistic Acoustic Environment in Cabs of Heavy-Duty Vehicles (OCTAVE)" under Grant No. P2024-01011.

\bibliographystyle{unsrturl}
\bibliography{references}

\end{document}